\documentclass[reprint,aip,jcp]{revtex4-1}

\usepackage{graphicx}
\usepackage{amssymb}
\usepackage{amsmath}
\usepackage{ifthen}
\usepackage{bbm}
\usepackage{xspace}
\usepackage[note-name=, use-sort-key = false]{notes2bib}
\usepackage{dcolumn}
\usepackage{longtable}

\newcommand{\Ml}[0]{$M_\Lambda$\xspace}
\newcommand{\Eh}[0]{$E_{\rm h}$\xspace}
\newcommand{\kcal}[0]{kcal mol$^{-1}$\xspace}

\newcommand{\Phiref}[0]{\tilde{\Phi}}

\newcommand{\cop}[1]{\hat{a}^{\dagger}_{#1}}
\newcommand{\aop}[1]{\hat{a}_{#1}}

\newcommand{\aphystei}[2]{\bra{#1}\!\!\ket{#2}}

\newcommand{\bra}[2][0]
{\ifthenelse{\equal{#1}{0}}{\left\langle #2 \right|}
{\ifthenelse{\equal{#1}{1}}{\big\langle #2 \big|}
{\ifthenelse{\equal{#1}{2}}{\Big\langle #2 \Big|}
{\ifthenelse{\equal{#1}{3}}{\bigg\langle #2 \bigg|}
{\ifthenelse{\equal{#1}{4}}{\Bigg\langle #2 \Bigg|}
{Error}}}}}
}

\newcommand{\bracket}[4][0]
{\ifthenelse{\equal{#1}{0}}{\left\langle #2 \middle| #3 \middle| #4 \right\rangle}
{\ifthenelse{\equal{#1}{1}}{\big\langle #2 \big| #3 \big| #4 \big\rangle}
{\ifthenelse{\equal{#1}{2}}{\Big\langle #2 \Big| #3 \Big| #4 \Big\rangle}
{\ifthenelse{\equal{#1}{3}}{\bigg\langle #2 \bigg| #3 \bigg| #4 \bigg\rangle}
{\ifthenelse{\equal{#1}{4}}{\Bigg\langle #2 \Bigg| #3 \Bigg| #4 \Bigg\rangle}
{Error}}}}}
}

\newcommand{\braket}[3][0]
{\ifthenelse{\equal{#1}{0}}{\left\langle #2 \middle| #3 \right\rangle}
{\ifthenelse{\equal{#1}{1}}{\big\langle #2 \big| #3 \big\rangle}
{\ifthenelse{\equal{#1}{2}}{\Big\langle #2 \Big| #3 \Big\rangle}
{\ifthenelse{\equal{#1}{3}}{\bigg\langle #2 \bigg| #3 \bigg\rangle}
{\ifthenelse{\equal{#1}{4}}{\Bigg\langle #2 \Bigg| #3 \Bigg\rangle}
{Error}}}}}
}

\newcommand{\ket}[2][0]
{\ifthenelse{\equal{#1}{0}}{\left| #2 \right\rangle}
{\ifthenelse{\equal{#1}{1}}{\big| #2 \big\rangle}
{\ifthenelse{\equal{#1}{2}}{\Big| #2 \Big\rangle}
{\ifthenelse{\equal{#1}{3}}{\bigg| #2 \bigg\rangle}
{\ifthenelse{\equal{#1}{4}}{\Bigg| #2 \Bigg\rangle}
{Error}}}}}
}

\begin{document}

%
%
\title{Adaptive multiconfigurational wave functions}

%
%
\author{Francesco A. Evangelista}
\email{francesco.evangelista@emory.edu}
\affiliation{Department of Chemistry and Cherry L. Emerson Center for Scientific Computation, Emory University, Atlanta, Georgia, USA}

%
%
\date{\today}

%
%
\begin{abstract}
A method is suggested to build simple multiconfigurational wave functions specified uniquely by an energy cutoff $\Lambda$.
These are constructed from a model space containing determinants with energy relative to that of the most stable determinant no greater than $\Lambda$.
The resulting $\Lambda$-CI wave function is adaptive, being able to represent both single-reference and multireference electronic states.
We also consider a more compact wave function parameterization ($\Lambda$+SD-CI), which is based on a small $\Lambda$-CI reference and adds a selection of all the singly and doubly excited determinants generated from it.
We report two heuristic algorithms to build $\Lambda$-CI wave functions.
The first is based on an approximate prescreening of the full configuration interaction space, while the second performs a breadth-first search coupled with pruning.
The  $\Lambda$-CI and $\Lambda$+SD-CI approaches are used to compute the dissociation curve of N$_2$ and the potential energy curves for the first three singlet  states of C$_2$.
Special attention is paid to the issue of energy discontinuities caused by changes in the size of the $\Lambda$-CI wave function along the potential energy curve.
This problem is shown to be solvable by smoothing the matrix elements of the Hamiltonian.
Our last example, involving the Cu$_2$O$_2^{2+}$ core, illustrates an alternative use of the $\Lambda$-CI method: as a tool to both estimate the multireference character of a wave function and to create a compact model space to be used in subsequent high-level multireference coupled cluster computations.
\end{abstract}

%
%
\maketitle
\section{Introduction}
One of the major obstacles to the numerical solution of the electronic Schr\"{o}dinger equation for systems of chemical interest is the factorial growth of the space of electronic configurations (or Slater determinants) with respect to the number of electrons and orbitals.
Conventional electronic structure approaches reduce the cost of solving the Schr\"{o}dinger equation from factorial to polynomial by means of compact and structured wave functions.
For example, single-reference coupled cluster (CC) theory expresses the wave function in terms of the exponential of a product of relatively few excitation operators.\cite{Crawford:2000ub,Bartlett:2007kv}
Likewise, complete-active-space (CAS) methods\cite{Szalay:2011uz} consider only electronic configurations generated by distributing a chosen number of electrons in a subset of the molecular orbitals.

However, with structure also comes rigidity.  Single-reference coupled cluster theory cannot properly describe multireference electronic states, as it inherently assumes that the wave function is dominated by a single Slater determinant.  CAS and more general active-space methods are also problematic.  The choice of the active space, often guided by chemical intuition, may be viewed as having a degree of arbitrariness, and cases have been discovered in which increasing the size of the active space does not immediately improve the accuracy of potential energy surfaces.\cite{Bauschlicher:1988be}
Moreover, a CAS wave function can only capture the static component of the correlation energy and must be augmented with a multireference treatment of dynamic correlation.

Maintaining a consistent definition of the active space while changing the molecular geometry is perhaps one of the most problematic issues of CAS methods.
Ideally, the active orbitals would lie in an energetic window well separated from the doubly occupied and virtual orbitals.  However, in realistic applications it is almost inevitable that active orbitals become degenerate or even switch order with core or virtual orbitals.
Consequently, potential energy surfaces computed with active-space methods are prone to incurable discontinuities.
When active spaces are combined with the self-consistent-field optimization of the molecular orbitals, new problems arise.
These include: convergence to a local energy minimum, bistability of the solutions, and spatial symmetry breaking.\cite{McLean:1985ik,SanchezDeMeras:1990gd,Guihery:1997cm}

\begin{figure*}[t]
\begin{center}
\includegraphics[width=6.5in]{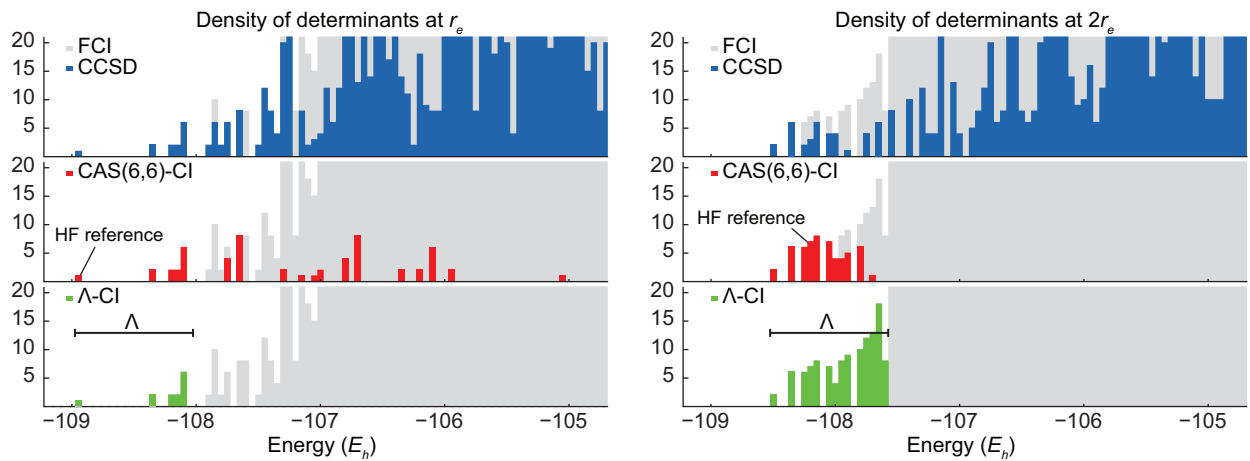}
\caption{\footnotesize  FCI density of determinants computed for N$_2$ at the equilibrium ($r_e$, where $r_e$ = 1.09768 \AA, left panel)  and stretched  (2$r_e$, right panel) geometries using the cc-pVDZ basis set and restricted-Hartree--Fock orbitals (in gray).  In addition, we show the space of determinants spanned by linearized CCSD (top, blue), CAS(6,6)-CI (middle, red), and $\Lambda$-CI (bottom, green) wave functions.}
\label{fig:n2-1re-dod}
\end{center}
\end{figure*}

We contend that the problems affecting coupled cluster theory and active space methods are caused by an unbalanced selection of the space of electronic configurations, which is a consequence of the rigid parameterization these approaches impose onto the wave function.
To illustrate this point, we will consider the \textit{density of determinants} for the full configuration interaction (FCI)  and approximate wave functions.
For a given wave function $\Psi$, we define the density of determinants as the histogram of the energies $E_I = \bra{\Phi_I} \hat{H} \ket{\Phi_I}$, where $\Phi_I$ is a generic Slater determinant contained in $\Psi$.
\bibnote{In the case of the FCI wave function, Gershgorin's circle theorem can be used to interpret the density of determinants as an approximation of the exact eigenvalue spectrum of the Hamiltonian.}
Fig.~\ref{fig:n2-1re-dod} shows the density of determinants of N$_2$ computed at the equilibrium ($r_e$) and stretched (2$r_e$) geometries using the FCI space, the linear component of the  CC with single and double excitations (CCSD) wave function, and the CAS-configuration interaction with six electrons distributed in six orbitals [CAS(6,6)-CI].
At the equilibrium distance, the CCSD wave function covers most of the low-energy determinants included in the FCI space, but it fails to account for low energetic triples and higher excitations.  This problem is even worse at the stretched geometry.
On the contrary, the CAS(6,6)-CI wave function captures the low-energy range of the FCI wave function at the stretched geometry, while at the equilibrium bond length it neglects important low energy excitations in favor of high energetic ones.
This example illustrates the difficulties encountered when designing a structured wave function required to model both the single- and multireference regimes of electron correlation.

The concept of adaptivity has been applied with success to generate optimal one-particle basis sets in self-consistent-field computations\cite{Laaksonen:1986gr,Becke:1990jq,Harrison:2004kt,Yamakawa:2005ke,Shiozaki:2007fz,Sekino:2008ef,Alizadegan:2010ky,Bischoff:2011kj} and second-order perturbation theory.\cite{Bischoff:2012fw}
In the case of many-body wave functions, the central issue is discerning which electronic configurations should enter a truncated CI or CC wave function.
The selection of configurations in CI has been developed long ago in a series of studies by Davidson and co-workers,\cite{Bender:1969va,Langhoff:1973wt} the MRD-CI method of Buenker and Peyerimhoff,\cite{Buenker:1974bl,Buenker:1978wd} and the CIPSI approach developed by Malrieu and co-workers which iteratively selects a CI space.\cite{Huron:1973cb,Evangelisti:1983hf}
These methods, and some more recent variants,\cite{Carter:1984gz,Cimiraglia:1987eg,Cave:1992ce,Miralles:1993gm,Steiner:1994el,Mitrushenkov:1995uk,Wenzel:1996kn,Angeli:1997kg,Angeli:1998hr,Hanrath:1997ch,Engels:2001vp,Nakatsuji:2005ba,Bunge:2006cl,Bytautas:2009hm,Roth:2009dm,Bytautas:2011eo,Sambataro:2011dj,LiManni:kr,LiManni:2013jm,Giner:2013tu,Alcoba:2013ci} use refined estimates of the importance of a configuration, which are obtained either from perturbation theory or by solving a small CI (see the review by Sherrill and Schaefer, Ref.~\onlinecite{Sherrill:1999tg}, for a in-depth analysis of these approaches).
All these methods are essentially adaptive, and in most cases are specified uniquely by a determinant (configuration) selection parameter.
An adaptive method that iteratively constructs a wave function in terms of nonorthogonal Slater determinants has been suggested by Koch and Dalgaard.\cite{Koch:1993tf}  More recently, Rodr\'{i}guez-Guzm\'{a}n and co-workers have generalized this method using instead projected Hartree--Fock determinants.\cite{RodriguezGuzman:2013bw}

The selection of determinants introduces a series of problems, including: discontinuities in the potential energy surface, lack of size consistency, and lack of orbital invariance---which implies a more pronounced dependence on the definition of the molecular orbital basis.
However, selected CI wave functions can be systematically improved, and various methods have been suggested to cure these deficiencies.
Discontinuities may be addressed by taking the union of CI wave functions at various geometries,\cite{Banerjee:1977bi} while the problem of size consistency has been addressed by Malrieu and co-workers with the creation of the size-consistent self-consistent CI approach.\cite{Daudey:1993ji,Meller:1994ey}
A self-consistent definition of the molecular orbitals has been suggested by Davidson,\cite{Bender:1969va} and uses an iterative determination of the natural orbitals for a selected CI.
Frozen natural orbitals,\cite{Barr:1970ec} are also known to accelerate the convergence of selected CI wave functions.
In addition, a number of specialized selection schemes that are optimal for the computation of dissociation energies and excitation energies have been proposed, including the correlation-consistent and dissociation-consistent CI approaches,\cite{Carter:1987ca,Carter:1988io} the difference-dedicated CI,\cite{Miralles:1993gm} aimed selection,\cite{Angeli:1997fs} multireference second-order perturbation theory with a selected reference,\cite{Grimme:2000gn} and the spectroscopic-oriented CI method.\cite{Neese:2003jg}

Progress has also been made in incorporating the idea of adaptivity in coupled cluster wave functions by selecting excitation operators.\cite{Nakatsuji:1991jt,Abrams:2005ui,Lyakh:2010ke}
Abrams and Sherrill\cite{Abrams:2005ui} have demonstrated that accurate and robust compact CC wave functions can be formed by selecting the configurations with the highest weights.
Lyakh and Bartlett\cite{Lyakh:2010ke} have formulated a scheme that automatically selects the most important amplitudes that enter a CC computation.
Stochastic sampling of the CC excitation space has been proposed by Thom.\cite{Thom:2010ua}
More recently,  Shen and co-workers\cite{Shen:2010es} and Melnichuk and Bartlett\cite{Melnichuk:2012kr} have applied  automatic schemes to select appropriate active orbitals to be used in active-space CC theories, and Landau \textit{et al.}\cite{Landau:2010hz} used natural occupation numbers to truncate the excitation space in the equation-of-motion CC method for ionized states.
It is also important to mention the growing interest in stochastic methods that sample the wave functions in the space of excited determinants.\cite{Gyorffy:2008ba,Ohtsuka:2008fk,Booth:2009hb,Petruzielo:2012ha,Tenno:2013kd,Coe:2013et}

In this work we propose a new class of multiconfigurational wave functions that can be used to represent multireference electronic states.
Our focus is to define a family of systematically improvable and adaptive wave functions that can properly describe the static component of electron correlation.
We envision using these adaptive wave functions to diagonalize an effective Hamiltonian obtained by a similarity transformation of the bare Hamiltonian, for example, the coupled cluster Hamiltonian or the effective Hamiltonian obtained by a unitary canonical transformation.\cite{Bartlett:2007kv,Yanai:2006gi}
However, in this work all examples are based on the bare Hamiltonian---which may be argued---is a more difficult test for our methods.

We work under the assumption that the static correlation energy is an intensive quantity.
This is the case when, for example, the wave function acquires multireference character as a consequence of breaking a small number of bonds or a localized electronic excitation.
Consequently, we will assume that the subset of the FCI space necessary to describe static electron correlation is small and does not grow with the size of the system.
This domain is somewhat complementary to that of the density-matrix renormalization group approach and other related methods suited to studying extended strongly correlated electron systems.\cite{White:1992tg,Zgid:2011fy,Knizia:2012do,Parker:2013ep}

Under these assumptions, an intriguing solution to the problem of choosing an active space is to consider zeroth-order wave functions defined by an energy cutoff parameter $\Lambda$.
Our idea is illustrated in the lower panels of Fig.~\ref{fig:n2-1re-dod}.
The bottom left panel shows that in the single-reference case, the wave function specified by $\Lambda$ contains only the Hartree--Fock reference and a few low-lying determinants.
In the multireference case (bottom right panel, Fig.~\ref{fig:n2-1re-dod}) the energy-based wave function increases in size and may be designed to include all the determinants accounted for by the CAS(6,6)-CI method.
Horoi, Brown, and Zelevinsky,\cite{Horoi:1994dk} have provided an interesting justification for the use of a energy-selected wave function in the context of nuclear structure computations.
Notice that the SplitCAS and SplitGAS methods of Li Manni and co-workers\cite{LiManni:kr,LiManni:2013jm} also use the energy as a selection criterion, but start from a CAS or a general active space (GAS).
The strategy outlined here can build an energy-selected model space without requiring the definition of a set of active orbitals, and thus it has the noteworthy property that the orbitals play a secondary role in determining the structure of the wave functions.

When applied to a CASCI wave function, the energy selection criterion can significantly reduce its size, because in general a large fraction of the determinants does not contribute to the wave function.
However, the energy-based selection criterion cannot screen for determinants that have a negligible coupling to the state that we are trying to represent.  This implies that the $\Lambda$-CI is not as efficient as the more sophisticated criteria used in the selected CI methods.
To address this problem, we also introduce a more efficient adaptive approach that combines the $\Lambda$-CI wave function with the selection criteria used in the MRD-CI and CIPSI methods.\cite{Bender:1969va,Langhoff:1973wt,Buenker:1974bl,Buenker:1978wd,Huron:1973cb,Evangelisti:1983hf}
More specifically, we use the energy criterion to automate the selection of a small set of reference determinants, which is then used to generate a selected multireference CI wave function.

To simplify our study we assume that the orbitals come from a restricted-Hartree--Fock calculation and are kept fixed.  By avoiding the orbital optimization process we remove the most problematic aspect of the CASSCF method, the nonlinear optimization process.
In principle, a more compact representation of the wave function can be achieved by resorting to natural orbitals,\cite{Bender:1969va} frozen natural orbitals,\cite{Barr:1970ec,Sosa:1989wz,Taube:2008ix} or improved virtual orbitals.\cite{Huzinaga:1970ky,Morokuma:1972fa}
These orbital choices can be easily combined with the $\Lambda$-CI method to improve its efficiency.

The practical realization of an energy-based adaptive strategy requires the development of technologies to identify a subspace of the FCI wave function without having to explicitly evaluate the energy of each of its elements.
Therefore we will describe two complimentary methods that allow us to identify Slater determinants with a relative energy (with respect to the lowest energy determinant) smaller than a given cutoff $\Lambda$.
We will consider two possible uses of the adaptive model space.
First, we will demonstrate that they are flexible zeroth-order wave functions capable of describing both single-reference and multireference electronic states.
Second, we will show that adaptive wave functions can be used to qualitatively analyze the electronic structure of transition metal compounds like Cu$_2$O$_2^{2+}$, whose FCI space contains about $10^{18}$ determinants.
We conclude the article with a discussion of the scaling properties of the adaptive model space and how it can be combined with a treatment of dynamical electron correlation.

%
%
\section{Theory}
\subsection{Definition of the energy-based adaptive model space \Ml}
Consider a system containing a given number of alpha and beta electrons ($N_\alpha$ and $N_\beta$, respectively) that occupy a set of $2K$ orthonormal spin orbitals $\{\phi_p\}$.
From this basis we can construct the space of Slater determinants that spans the full configuration interaction wave function,  $M_{\rm FCI} = \{\Phi_I\}$, of dimension $N_{\rm FCI} = \dim M_{\rm FCI}$.
We assume that the determinants in $M_{\rm FCI}$ are sorted according to the expectation value of the Hamiltonian operator $E_I = \bra{\Phi_I} \hat{H} \ket{\Phi_I}$, so that $E_{I} \leq E_{I+1}$.
In our notation, the index $I$ runs from 0 to $N_{\rm FCI}-1$, and thus $E_0 = \bra{\Phi_0} \hat{H} \ket{\Phi_0}$ is the energy corresponding to the determinant with the lowest energy.

Given an energy cutoff $\Lambda$, we define the adaptive model space ($M_{\Lambda}$) as the set of determinants with relative energy $E_I - E_0$ less or equal to $\Lambda$:
\begin{equation} \label{eq:lambda_model_space}
M_\Lambda = \{ \ket{\Phi_I} :  E_I - E_{0} \leq \Lambda \}.
\end{equation}
Defining the model space in terms of the energy of Slater determinants instead of configuration state functions simplifies the formulation of the algorithms used to construct \Ml.\bibnote{It is important to point out that a model space defined in terms of the energy of configuration state functions would not be identical to the one considered in this work and based on the energy of a determinant.}
However, the tradeoff of this choice is that \Ml spans eigenfunctions of $\hat{H}$ with different value of total spin.
The set of determinants in \Ml form a basis for the $\Lambda$-configuration interaction ($\Lambda$-CI) wave function:
\begin{equation}
\ket{\Psi_{\Lambda\text{-CI}}} = \sum_{I \in M_\Lambda} C^{\Lambda}_I \ket{\Phi_I},
\end{equation}
where the coefficients $C_I^\Lambda$ and the energy of the $\Lambda$-CI wave function ($E_{\Lambda}$) are obtained by solving the eigenvalue equation:
\begin{equation}
\hat{H} \ket{\Psi_{\Lambda\text{-CI}}} = E_{\Lambda} \ket{\Psi_{\Lambda\text{-CI}}}.
\end{equation}

Because a selection criterion based uniquely on the energy is not highly efficient, we also consider a combination of the $\Lambda$-CI wave function with the selection schemes used in the MRD-CI and CIPSI methods.\cite{Bender:1969va,Langhoff:1973wt,Buenker:1974bl,Buenker:1978wd,Huron:1973cb,Evangelisti:1983hf}
To this end we assume $\Lambda$ to be a sufficiently small energy threshold (about 1--2 \Eh) so that the  corresponding $\Lambda$-CI wave function ($\Psi_{\Lambda\text{-CI}}$) includes a manageable number of determinants.
We then augment this space with the set $M_{\Lambda}^{\rm SD}$, which contains part of all the singly and doubly excited determinants formed out of \Ml.  This space is selected to include the determinants that are estimated to give the largest contribution to the energy or the wave function.
The determinants from the sets \Ml and $M_{\Lambda}^{\rm SD}$ are used to build the $\Lambda$+SD-CI wave function:
\begin{equation}
\ket{\Psi_{\Lambda\text{+SD-CI}}}
= \sum_{\Phi_I \in M_\Lambda} C^{\Lambda\rm +SD}_I \ket{\Phi_I}
+ \sum_{\Phi_A \in M_{\Lambda}^{\rm SD}} C^{\Lambda\rm +SD}_A \ket{\Phi_A}.
\end{equation}
The $\Lambda$+SD-CI wave function is essentially a selected uncontracted multireference CI wave function, like in the MRD-CI approach, but with the CAS reference replaced by a $\Lambda$-CI wave function.
Since the selection step performed after the computation of the $\Lambda$-CI wave function can \textit{a priori} discard excitations to Rydberg or charge-transfer states that give negligible contributions to the energy and to the wave function, the $\Lambda$+SD-CI approach is expected to be applicable to larger systems and to be more efficient that the $\Lambda$-CI scheme.

In this work we consider four approaches to form $M_{\Lambda}^{\rm SD}$.
The first method selects a determinant according to an estimate of its second-order perturbation theory energy contribution [$\epsilon_I^{(2)}$] assuming the Epstein-Nesbet partitioning of $\hat{H}$:
\begin{equation} \label{eq:mp2_screening}
\epsilon_I^{(2)} = \frac{|\bra{\Psi_{\Lambda}} \hat{H} \ket{\Phi_I}|^2}{E_{\Lambda} - E_I},
\end{equation}
and includes $\Phi_I$ in $M_{\Lambda}^{\rm SD}$ if $|\epsilon_I^{(2)}|$ is greater or equal than the threshold $\tau_\epsilon$.
This selection scheme has been advocated by Davidson and co-workers,\cite{Bender:1969va,Langhoff:1973wt} and it is used in the MRD-CI approach of Buenker and Peyerimhoff.\cite{Buenker:1974bl,Buenker:1978wd}
Notice, that by summing up all the energy estimates of the discarded determinants it is possible to obtain a second-order  energy correction to the selected CI energy:
\begin{equation} \label{eq:EN2correction}
E^{(2)}_{\rm corr} = \sum_{\Phi_I \notin M_{\Lambda}^{\rm SD}} \epsilon_I^{(2)}.
\end{equation}
This estimate can be used to correct the $\Lambda$+SD-CI energy for the singles and doubles not included in $M_{\Lambda}^{\rm SD}$.

The second selection scheme, used by Huron, Malrieu, and Rancurel in CIPSI\cite{Huron:1973cb} and later by others,\cite{Mitrushenkov:1995uk,Roth:2009dm} is based on a first-order perturbation theory (PT) estimate of the the CI coefficient assuming the Epstein-Nesbet partitioning of the Hamiltonian.
In this selection scheme, a determinant $\Phi_I$ is included in $M_{\Lambda}^{\rm SD}$ if the absolute value of the first-order PT coefficient [$C_I^{(1)}$] is greater or equal than a threshold $\tau_C$:
\begin{equation} \label{eq:roth_screening}
|C_I^{(1)}| = \left|\frac{\bra{\Psi_{\Lambda}} \hat{H} \ket{\Phi_I}}{E_{\Lambda} - E_I}\right|
= \left|\frac{\sum_{J \in M_{\Lambda}}C^{\Lambda}_J \bra{\Phi_J} \hat{H} \ket{\Phi_I}}{E_{\Lambda} - E_I}\right|
\geq \tau_C.
\end{equation}
Notice, that Eqs.~\eqref{eq:EN2correction}-\eqref{eq:roth_screening} are evaluated using $\Psi_{\Lambda}$ and $E_\Lambda$ from the initial $\Lambda$-CI computation.

Following Angeli and Persico,\cite{Angeli:1997fs} we also consider an \textit{aimed} version of these two selection criteria.
For example, in the aimed variant based on $\epsilon_I^{(2)}$, we sort the determinants according to the value of $|\epsilon_I^{(2)}|$ (in descending order), and include a given number of the sorted determinants (indicated with the symbol $N_{\rm CI}$) in the set $M_{\Lambda}^{\rm SD}$, ensuring that the sum of $|\epsilon_I^{(2)}|$ ($\sigma_\epsilon$) for the determinants that are excluded from $M_{\Lambda}^{\rm SD}$ is less than the parameter $\tau_\epsilon$:
\begin{equation}
\sigma_\epsilon = \sum_{\Phi_I \notin M_{\Lambda}^{\rm SD}} |\epsilon_I^{(2)}| < \tau_\epsilon.
\end{equation}
An analogous procedure is used in the aimed selection of the determinants according to the value of $C_I^{(1)}$, and the space $M_{\Lambda}^{\rm SD}$ is selected to satisfy: 
\begin{equation}
\sigma_C = \sum_{\Phi_I \notin M_{\Lambda}^{\rm SD}} |C_I^{(1)}|^2 < \tau_C,
\end{equation}
where $\tau_C$ is a selection parameter.
Thus, the aimed schemes try to generate results with a guaranteed bound on the error in the energy or the wave function.

The $\Lambda$+SD-CI method can be generalized to treat several electronic states of the same symmetry at a time. Here we consider only the case of threshold-based criteria and generate a \textit{single} model space optimized for all the target states.
In this case we generalize the selection criteria by computing $\epsilon_I^{(2)}$ or $C_I^{(1)}$ for all the excited states under consideration.  A determinant $\Phi_I$ is included in $M_{\Lambda}^{\rm SD}$ if the largest of the values of $\epsilon_I^{(2)}$ or $C_I^{(1)}$ obtained for all the excited states is greater than or equal to $\tau_\epsilon$ or $\tau_C$.

\subsection{A prescreening algorithm for constructing \Ml} \label{sec:algorithm_for_M_Lambda}
Constructing the set \Ml appears to be a nontrivial task.
Recall that the energy expression for a generic Slater determinant $\Phi_I$ is a quadratic function of the occupation numbers of each spin orbital $\phi_p$ ($n_p^I =\{0,1\}$), the one-electron integrals $h_{pq} = \bra{\phi_p} \hat{h} \ket{\phi_q}$, and the diagonal part of the antisymmetrized two-electron integrals $V_{pq} = \aphystei{pq}{pq}$:
\begin{equation}\label{eq:det_energy}
E_I =  \sum_p h_{pp} \, n_p^I+ \frac{1}{2} \sum_{pq}  V_{pq} \,n_p^I \,n_q^I.
\end{equation}
Even finding the determinant that minimizes the energy---which is a simpler question equivalent to a binary quadratic programming problem---may necessitate an exhaustive search trough the entire space of configurations.
Determining \Ml requires going one step further: finding the subset of constrained binary vectors such that the value of the quadratic form [Eq.~\eqref{eq:det_energy}]  is less than $\Lambda$.
It is unlikely that one can formulate a procedure that will directly construct the set \Ml.
Therefore, we have formulated two heuristic algorithms that can accomplish this task.
These are not guaranteed to find all the determinants in \Ml, but can be applied simultaneously as a consistency test.

The strategy that we adopt in the first algorithm is prescreening the FCI space using a linear approximation of Eq.~\eqref{eq:det_energy}.
We first assume to have found a determinant $\Phiref$ with energy $\tilde{E}$ that is reasonably close to the lowest energy determinant $\Phi_0$.
A generic determinant $\Phi_I$ may then be expressed in terms of an excitation operator ($\hat{a}_{ij\cdots}^{ab\cdots}$) acting on the reference $\Phiref$:
\begin{equation} \label{eq:excitation}
\ket{\Phi_I} = \hat{a}_{ij\cdots}^{ab\cdots} \ket[1]{\Phiref} = \cop{a}  \cop{b} \cdots  \aop{j} \aop{i} \ket[1]{\Phiref},
\end{equation}
where $\cop{}$ and $\aop{}$ are second quantization creation and annihilation operators, respectively, and the indices $i,j,\ldots$ and $a,b,\ldots$ represent the occupied and unoccupied orbitals of $\Phiref$, respectively.

We then write the energy difference $E_I - \tilde{E}$ as a quadratic function of the \textit{difference} of the occupation numbers of $\Phi_I$ and $\Phiref$ ($\delta n^I_p = n^I_p - \tilde{n}_p$ ):
\begin{equation} 
\label{eq:det_energy_difference}
E_I - \tilde{E} =\sum_p \tilde{\epsilon}_p \, \delta n^I_p  + \frac{1}{2} \sum_{pq} V_{pq}
\, \delta n^I_p  \, \delta n^I_q,
\end{equation}
where
\begin{equation} \label{eq:epsilon}
\tilde{\epsilon}_p = h_{pp}  + \sum_q \aphystei{pq}{pq} \tilde{n}_q,
\end{equation}
is the Hartree--Fock orbital energy computed using the occupation numbers of $\Phiref$.
Neglecting the quadratic term in Eq.~\eqref{eq:det_energy_difference} we obtain a formula to estimate the relative energy  of $\Phi_I$:
\begin{equation} \label{eq:partitioned_energy_difference}
E_I - \tilde{E} \approx  
\Delta_{ij\cdots}^{ab\cdots}  =
\Delta^{ab\cdots} - \Delta_{ij\cdots}
\equiv \tilde{\epsilon}_a + \tilde{\epsilon}_b + \cdots -\tilde{\epsilon}_i - \tilde{\epsilon}_j - \cdots,
\end{equation}
where we introduce the M{\o}ller--Plesset energy denominator ($\Delta_{ij\cdots}^{ab\cdots}$) and partition the sum into the creation ($\Delta^{ab\cdots}$) and annihilation ($\Delta_{ij\cdots}$) components of $\hat{a}_{ij\cdots}^{ab\cdots}$.

Eq.~\eqref{eq:partitioned_energy_difference} allows us to form a set of trial determinants, $M_{\Lambda'}^{\rm trial}$:
\begin{equation}
M_{\Lambda'}^{\rm trial} = \{\hat{a}_{ij\cdots}^{ab\cdots}\ket[1]{\tilde{\Phi}} : \Delta_{ij\cdots}^{ab\cdots}  \leq \Lambda' \},
\end{equation}
with the threshold $\Lambda'$ assumed to be looser than $\Lambda$ ($\Lambda' > \Lambda$).
By choosing an appropriate value of $\Lambda'$, $M_{\Lambda'}^{\rm trial}$ can be made large enough to contain all the elements of \Ml.

Our algorithm based on prescreening visits each element of the trial model space, $\Phi^{\rm trial}_I \in M_{\Lambda'}^{\rm trial}$, computes its energy $E_I^{\rm trial} = \bra{\Phi^{\rm trial}_I}\hat{H}\ket{\Phi^{\rm trial}_I}$, and determines if it belongs to \Ml.
It is important to notice that our algorithm does not require to store $M_{\Lambda'}^{\rm trial}$.
To facilitate the construction of $M_{\Lambda'}^{\rm trial}$, we form prescreened lists of annihilation ($\mathcal{A}_{\Lambda'}$) and creation ($\mathcal{C}_{\Lambda'}$) operators:
\begin{align}
\mathcal{A}_{\Lambda'} &= \{ (\Delta_{ij\cdots},\aop{i}\aop{j}\cdots) : -\Delta_{ij\cdots} \leq \Lambda' \}, \\
\mathcal{C}_{\Lambda'} &= \{ (\Delta^{ab\cdots},\cop{a}\cop{b}\cdots) : \Delta^{ab\cdots} \leq \Lambda' \}.
\end{align}
These lists store pairs of M{\o}ller--Plesset denominators and their corresponding operators and are sorted according the denominator energy.
The critical aspect of the prescreening algorithm is that the lists $\mathcal{A}_{\Lambda'}$ and $\mathcal{C}_{\Lambda'}$ can be generated very efficiently, and that they permit to construct $M_{\Lambda'}^{\rm trial}$ directly by combining strings that satisfy $\Delta_{ij\cdots}^{ab\cdots}  \leq \Lambda'$.

This algorithm does not assume to know the determinant with minimum energy. 
Thus, during the screening $E_I^{\rm trial}$ is compared to an approximate value of $E_0$ that we indicate with $E_{\rm min}$ and is initialized with $\tilde{E}$.
At the same time, if a determinant with energy lower than $E_{\rm min}$ is found, the value of $E_{\rm min}$ is updated.
In this way, our algorithm also performs a search for the determinant with the lowest energy.
At the end of its execution $E_0 = E_{\rm min}$.
Therefore, after visiting $M_{\Lambda'}^{\rm trial}$ it is necessary to revisit all of the determinants that are included in \Ml, and eliminate those that do not satisfy the selection criterion: $E_I - E_0 \leq \Lambda$.

An alternative algorithm for constructing \Ml is reported in Appendix A.

%
%
\section{Computational details}
The approximate prescreening algorithm to construct the set \Ml is implemented as plugin in the \textsc{Psi4} program package.\cite{Turney:2011gr}
Our code is a generalization of the algorithm presented here, which in addition to separately screening the occupied and virtual strings, also takes into account spin and spatial symmetry and the excitation level of the excitation operators.
Once the set \Ml is built (or \Ml and $M_{\Lambda}^{\rm SD}$, in the case of the $\Lambda$+SD-CI method), we form the Hamiltonian matrix $H_{IJ} = \bra{\Phi_I} \hat{H} \ket{\Phi_J}$ with $\Phi_I, \Phi_J \in M_{\Lambda}$ using Slater's rules, storing only those elements that are not equal to zero.
The Hamiltonian matrix is diagonalized using the Davidson--Liu algorithm\cite{Davidson:1975db,Liu:1978ve} to obtain the energy and coefficients of the CI wave function.
The most expensive part of our computations is the construction and diagonalization of the Hamiltonian matrix.  Both these steps have a computational cost that scales as $N_{\rm CI} \,K^4$, where $N_{\rm CI}$ is the number of determinants contained in the set \Ml or $M_{\Lambda} \cup M_{\Lambda}^{\rm SD}$.

All computations use restricted Hartree--Fock orbitals.
In mean-field computations at non-equilibrium distances, the number of occupied orbitals per irreducible representation is kept fixed at the optimal value for the equilibrium bond length.
Numerical results for N$_2$ and C$_2$ were computed using the 6-31G and the 6-31G* basis sets,\cite{Hehre:1972gg,Hariharan:1973gd} using Cartesian atomic orbitals.
All computations, except when explicitly mentioned, were carried out freezing the the 1s atomic-like orbitals of N and O, and the 1s--3p atomic-like orbitals of Cu.
Computations of the Cu$_2$O$_2^{2+}$ core employed the cc-pVTZ basis set and spherical atomic orbitals.\cite{Dunning:1989uk,Balabanov:2005gq}
Convergence of the Mk-MRCC equations was facilitated by Tikhonow regularization of the Mk-MRCC equations setting the shift parameter $\omega$ equal to 0.1 \Eh.\cite{Taube:2009jz,Das:2010bj}

%
%
\section{Results}

\begingroup
\squeezetable
\begin{table*}[tbp]
\centering
\caption{
Comparison of $\Lambda$-CI and $\Lambda$+SD-CI wave functions.  Ground singlet state of  N$_2$ computed at $r$(N-N) = $r_e$ and $2 r_e$, using the 6-31G basis set and restricted-Hartee--Fock orbitals; all electrons were included in the correlated wave functions.
Convergence of the total energy ($E$), the energy error with respect to FCI ($\Delta E$), the total number of determinants ($N_{\rm CI}$), the relative energy error at the two geometries (NPE), and the ratio between $N_{\rm CI}$ at $2r_e$ and $r_e$ ($N_{\rm CI}^{2r_e}$/$N_{\rm CI}^{r_e}$)  vs. $\tau$ or $\sigma$.}
\label{tab:n2-convergence}
\begin{tabular*}{6.5in}{@{\extracolsep{\fill}}cccr@{}rcr@{}rrr}
\hline\hline
& & \multicolumn{3}{c}{$r$(N-N) = $r_e$} & \multicolumn{3}{c}{$r$(N-N) = $2 r_e$}\\
\cline{3-5} \cline{6-8}
 $\Lambda$ & $\tau$/$\sigma$ & $E$ & $\Delta E$ & $N_{\rm CI}^{r_e}$ & $E$ & $\Delta E$ & $N_{\rm CI}^{2r_e}$ & NPE & $N_{\rm CI}^{2r_e}$/$N_{\rm CI}^{r_e}$ \\
(\Eh) &(\Eh) & (\Eh) &  (m\Eh) &  &  (\Eh) &  (m\Eh) &  & (m\Eh) & \\
\hline
\multicolumn{10}{c}{$\Lambda$-CI}
\\
0 & - & $-$108.867764 & 237.17 & 1 & $-$108.516412 & 333.27 & 1 & $-$96.10 & 1.00\\
1 & - & $-$108.941581 & 163.35 & 13 & $-$108.728715 & 120.96 & 154 & 42.39 & 11.85\\
2 & - & $-$108.995664 & 109.27 & 294 & $-$108.779191 & 70.49 & 2474 & 38.78 & 8.41\\
3 & - & $-$109.062715 & 42.22 & 2665 & $-$108.821554 & 28.12 & 18518 & 14.09 & 6.95\\
4 & - & $-$109.090184 & 14.75 & 15935 & $-$108.844135 & 5.54 & 87260 & 9.20 & 5.48\\
4.5 & - & $-$109.094444 & 10.49 & 32852 & $-$108.846105 & 3.57 & 163382 & 6.91 & 4.97\\
$\infty$ & - & $-$109.104933 & 0.00 & 126608256 & $-$108.849679 & 0.00 & 126608256 & 0.00 & 1.00\\
[3pt]\multicolumn{10}{c}{$\Lambda$+SD-CI ($|\epsilon_I^{(2)}| \geq \tau_\epsilon$)}
\\
2 & $1\times10^{-5}$ & $-$109.089750 & 15.18 & 1002 & $-$108.821193 & 28.49 & 3807 & $-$13.30 & 3.80\\
2 & $1\times10^{-6}$ & $-$109.097015 & 7.92 & 2754 & $-$108.840108 & 9.57 & 8491 & $-$1.65 & 3.08\\
2 & $1\times10^{-7}$ & $-$109.100339 & 4.59 & 8038 & $-$108.845485 & 4.19 & 20783 & 0.40 & 2.59\\
2 & $1\times10^{-8}$ & $-$109.101596 & 3.34 & 21341 & $-$108.847690 & 1.99 & 51148 & 1.35 & 2.40\\
2 & $1\times10^{-9}$ & $-$109.102123 & 2.81 & 43832 & $-$108.848245 & 1.43 & 109113 & 1.38 & 2.49\\
[3pt]\multicolumn{10}{c}{$\Lambda$+SD-CI ($|C_I^{(1)}| \geq \tau_C$)}
\\
2 & $1\times10^{-3}$ & $-$109.092275 & 12.66 & 1386 & $-$108.833985 & 15.69 & 5635 & $-$3.04 & 4.07\\
2 & $5\times10^{-4}$ & $-$109.096351 & 8.58 & 2638 & $-$108.840077 & 9.60 & 8637 & $-$1.02 & 3.27\\
2 & $1\times10^{-4}$ & $-$109.100847 & 4.09 & 11111 & $-$108.846273 & 3.41 & 26111 & 0.68 & 2.35\\
2 & $5\times10^{-5}$ & $-$109.101504 & 3.43 & 18476 & $-$108.847481 & 2.20 & 44392 & 1.23 & 2.40\\
2 & $1\times10^{-5}$ & $-$109.102191 & 2.74 & 42343 & $-$108.848300 & 1.38 & 111096 & 1.36 & 2.62\\
[3pt]\multicolumn{10}{c}{$\Lambda$+SD-CI ($\sigma_\epsilon < \tau_\epsilon$)}
\\
2 & $1\times10^{-2}$ & $-$109.087602 & 17.33 & 841 & $-$108.834635 & 15.04 & 5749 & 2.29 & 6.84\\
2 & $1\times10^{-3}$ & $-$109.099287 & 5.65 & 5276 & $-$108.846051 & 3.63 & 24328 & 2.02 & 4.61\\
2 & $1\times10^{-4}$ & $-$109.101568 & 3.36 & 20869 & $-$108.848076 & 1.60 & 77223 & 1.76 & 3.70\\
2 & $1\times10^{-5}$ & $-$109.102151 & 2.78 & 46102 & $-$108.848410 & 1.27 & 185793 & 1.51 & 4.03\\
[3pt]\multicolumn{10}{c}{$\Lambda$+SD-CI ($\sigma_C < \tau_C$)}
\\
2 & $1\times10^{-3}$ & $-$109.093165 & 11.77 & 1538 & $-$108.840195 & 9.48 & 8778 & 2.28 & 5.71\\
2 & $1\times10^{-4}$ & $-$109.100321 & 4.61 & 8052 & $-$108.846781 & 2.90 & 31350 & 1.71 & 3.89\\
2 & $1\times10^{-5}$ & $-$109.101714 & 3.22 & 22045 & $-$108.848119 & 1.56 & 79872 & 1.66 & 3.62\\
2 & $1\times10^{-6}$ & $-$109.102200 & 2.73 & 42878 & $-$108.848428 & 1.25 & 171428 & 1.48 & 4.00\\
[3pt]\multicolumn{10}{c}{$\Lambda$+SD-CI ($|\epsilon_I^{(2)}| \geq \tau_\epsilon$) + $E^{(2)}_{\rm corr}$} 
\\
2 & $1\times10^{-5}$ & $-$109.097608 & 7.32 & 1002 & $-$108.841983 & 7.70 & 3807 & $-$0.37 & 3.80\\
2 & $1\times10^{-6}$ & $-$109.099259 & 5.67 & 2754 & $-$108.845297 & 4.38 & 8491 & 1.29 & 3.08\\
2 & $1\times10^{-7}$ & $-$109.100894 & 4.04 & 8038 & $-$108.846781 & 2.90 & 20783 & 1.14 & 2.59\\
2 & $1\times10^{-8}$ & $-$109.101691 & 3.24 & 21341 & $-$108.847936 & 1.74 & 51148 & 1.50 & 2.40\\
2 & $1\times10^{-9}$ & $-$109.102135 & 2.80 & 43832 & $-$108.848290 & 1.39 & 109113 & 1.41 & 2.49\\
\hline\hline
\end{tabular*}
\end{table*}
\endgroup

\subsection{Analysis of the prescreening algorithm}
We first provide an analysis of the approximate prescreening method for building \Ml and show that if a sufficiently large value of $\Lambda'$ is chosen, then our algorithm will find all the elements of \Ml.
Our analysis will consider two cases: (a) the equilibrium ($r_e$ = 1.09768 \AA{}, from Ref.~\onlinecite{Herzberg:i0hQ0hVT}) and (b) stretched geometry ($2r_e$) of N$_2$, which correspond respectively to a single- and multireference wave function.

\begin{figure*}[t]
\begin{center}
\includegraphics[width=6.5in]{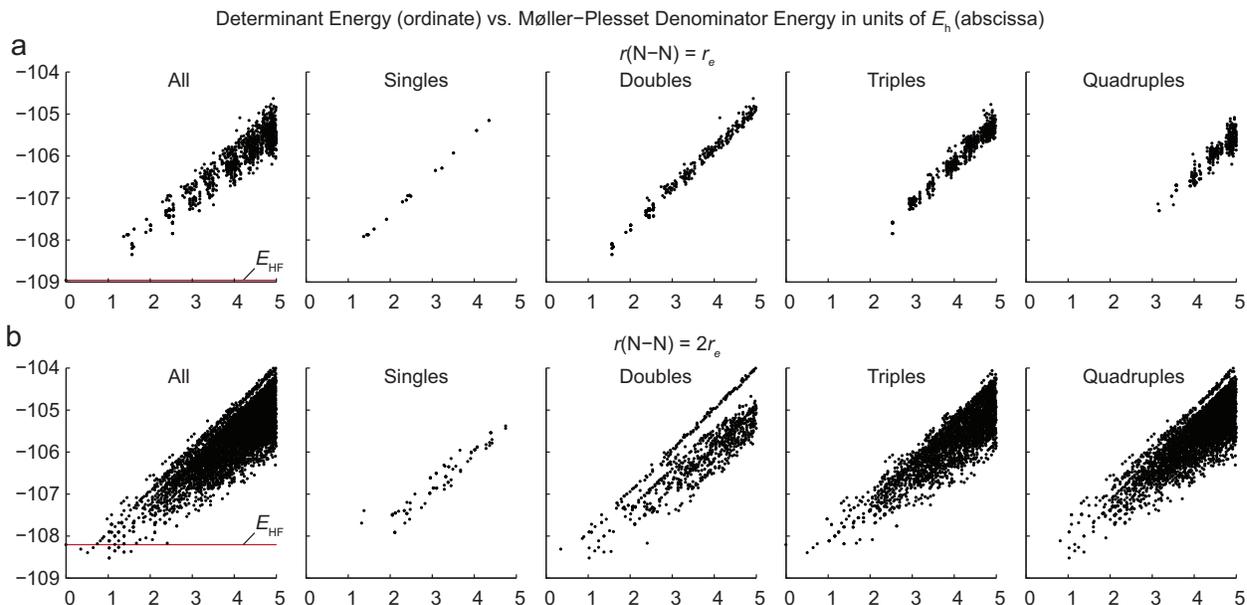}
\caption{Distribution determinant energies vs. M{\o}ller--Plesset denominators (in \Eh) for N$_2$ computed using RHF orbitals and the cc-pVDZ basis set.  The left-most plot shows the sum of the contributions from all excitation classes, while the remaining plots depict contributions from individual excitation levels.  Distribution computed at: (a) the equilibrium geometry, and (b) twice the equilibrium geometry.  The red horizontal line shows the energy of the Hartree--Fock determinant.}
\label{fig:n2-ddplot}
\end{center}
\end{figure*}

In Fig.~\ref{fig:n2-ddplot} we show the distribution of the determinant energies $E_I$ and the corresponding M{\o}ller--Plesset denominators $\Delta_{ij\cdots}^{ab\cdots}$ for the two geometries.
The approximate prescreening algorithm is guaranteed to work when a linear function of the denominator energy can be found that is a lower bound to the determinant energy:
\begin{equation} \label{eq:bound}
E_I > \alpha + \beta \Delta_{ij\cdots}^{ab\cdots},
\end{equation}
where $\beta > 0$ and $\Delta_{ij\cdots}^{ab\cdots}$ is the denominator corresponding to the excitation that generates $\Phi_I$ from $\tilde{\Phi}$ [see Eq.~\eqref{eq:excitation}].
If Eq.~\eqref{eq:bound} is satisfied, then by choosing a sufficiently large value of $\Lambda'$ we are guaranteed to be able to find all the determinants that fall within a given energy cutoff $\Lambda$.
For both the distributions shown in Fig.~\ref{fig:n2-ddplot} it is possible to satisfy the lower bound condition expressed by Eq.~\eqref{eq:bound}.
The lower panel of Fig.~\ref{fig:n2-ddplot} shows also an interesting feature: for elongated N$_2$ the  Hartree--Fock determinant is not the lowest in energy.
A few doubly-, triply-, and quadruply-excited determinants lie below it.

\subsection{Comparison of the various adaptive wave functions} \label{sec:comparison}
Next, the $\Lambda$-CI and $\Lambda$+SD-CI approaches are compared by computing the energy of N$_2$ at the geometries used in the previous section, using the 6-31G basis set and correlating all the electrons.
Table~\ref{tab:n2-convergence} reports the total energy, the energy error with respect to FCI, and the size of various $\Lambda$-CI wave function for values of $\Lambda$ in the range 1--4.5 \Eh.
The $\Lambda$-CI energy shows a consistent reduction of the error as $\Lambda$ is increased.
In the computation with the largest value of $\Lambda$, 4.5 \Eh, the $\Lambda$-CI wave function computed at the N$_2$ equilibrium geometry contains 32852 determinants, and the error with respect to FCI is ca. 10.5 m\Eh.
At the stretched geometry, the $\Lambda$-CI wave function contains 163382 determinants, and the energy deviates from the FCI value by ca. 3.6 m\Eh.
When compared to the size of the FCI space---126608256 determinants---these wave functions are very compact and they recover a large part of the correlation energy.
However, the NPE,\bibnote{The nonparallelism error (NPE) is commonly defined as the difference between the maximum and minim of the error with respect to FCI [$\Delta E(r)$] over a range of geometries $R$: $\mathrm{NPE} = \max_{r \in R}[\Delta E(r)] - \min_{r \in R}[\Delta E(r)]$} shown in the penultimate column of Table~\ref{tab:n2-convergence}, is found to be quite large: 6.9 m\Eh in the case of the $\Lambda$ = 4.5 \Eh wave function.

The $\Lambda$+SD-CI wave functions are found to be significantly more efficient than the $\Lambda$-CI ones.
This point is illustrated in Table~\ref{tab:n2-convergence} by taking a reference $\Lambda$-CI with $\Lambda$ = 2 \Eh, which at the equilibrium and stretched geometry contains respectively 294 and 2474 determinants.
At the equilibrium geometry, the $\Lambda$+SD-CI wave function obtained by neglecting all the determinants with $|\epsilon_I^{(2)}| < \tau_\epsilon$ = $10^{-6}$ \Eh, can achieve an error with respect to FCI of only 7.9 m\Eh, using just 2754 determinants.
This error should be compared to that of the reference $\Lambda$-CI wave function, which is about 109 m\Eh.
By reducing $\tau_\epsilon$ further to $10^{-9}$ \Eh, the $\Lambda$+SD-CI wave function grows to 48832 determinants, and the energy error is reduced to only 2.81 m\Eh and the NPE to ca. 1.38 m\Eh (ca. 0.9 \kcal).

Overall, determinant selection based on  $\epsilon_I^{(2)}$ and  $C_I^{(1)}$ yield very similar results.
For example,  setting $\tau_\epsilon = 10^{-9}$ \Eh, the former selection method yields a wave function that contains respectively (43832, 109113) determinants at ($r_e$, $2r_e$), and yields energies that differ from the FCI values by (2.81, 1.43) m\Eh.  The latter selection scheme with $\tau_C = 10^{-5}$, produces a similar number of determinants (42343,  111096) at ($r_e$, $2r_e$), and yields energies that differ from the FCI results by (2.74, 1.38) m\Eh.
The aimed selection schemes appear to be as efficient as the threshold-based selection.  However, the last column of Table~\ref{tab:n2-convergence} shows that the ratio between the size of the wave function at the stretched and equilibrium geometries is larger for the aimed methods, with no significant reduction in the energy error.

The bottom of table~\ref{tab:n2-convergence} shows the energy for the $|\epsilon_I^{(2)}|$-selected $\Lambda$+SD-CI method plus the second-order energy correction, $E^{(2)}_{\rm corr}$ [Eq.~\eqref{eq:EN2correction}]. 
This correction appears to be particularly useful when using a large selection threshold.  For example, in the case $\tau_\epsilon = 10^{-5}$, $E^{(2)}_{\rm corr}$ reduces the NPE of the $|\epsilon_I^{(2)}|$-selected $\Lambda$+SD-CI scheme from $-13.30$ to only $-0.37$ m\Eh.

\subsection{Dissociation curve of N$_2$} \label{sec:n2}
\begin{figure*}[t]
\begin{center}
\includegraphics[width=6.5in]{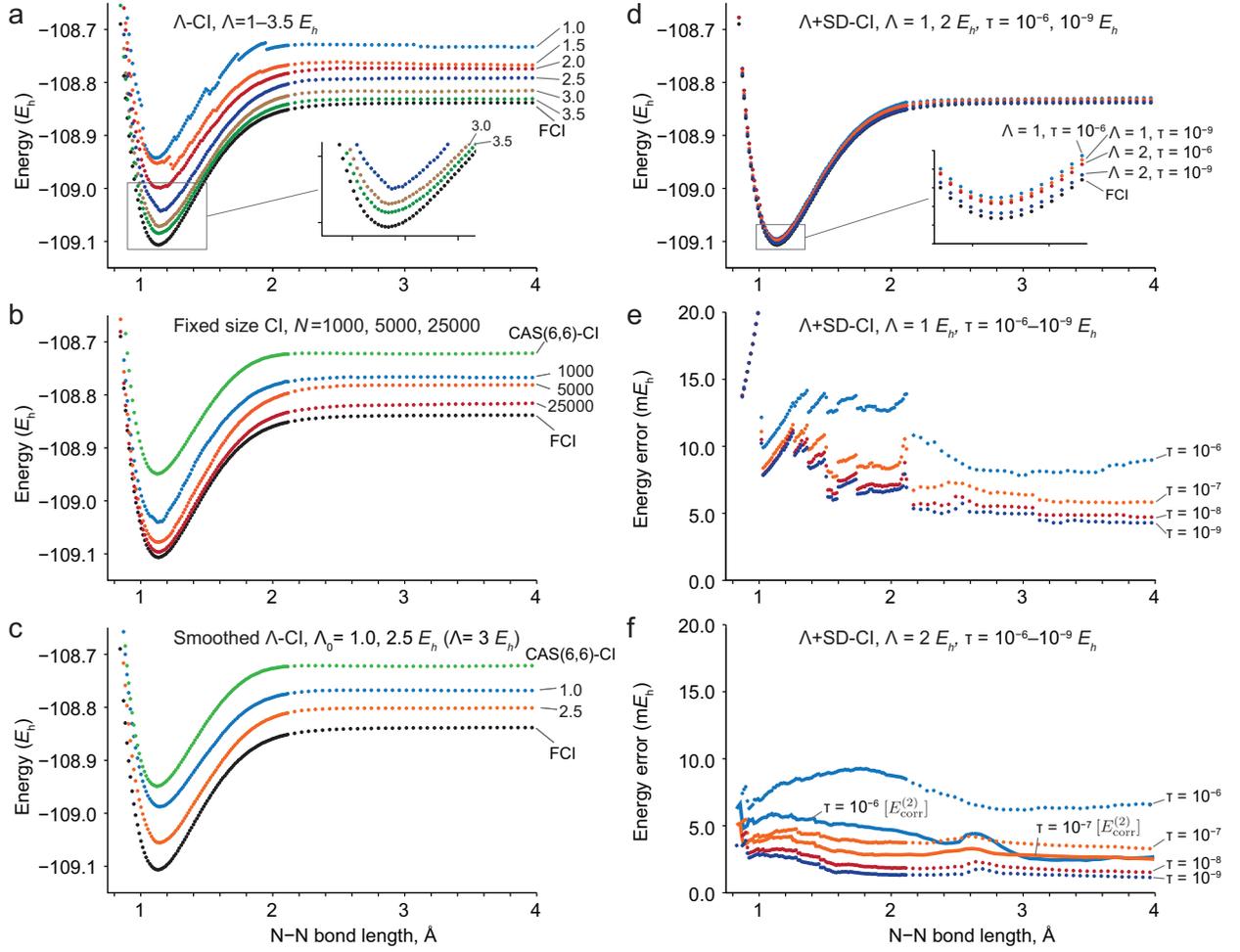}
\caption{$\Lambda$-CI and $\Lambda$+SD-CI potential energy curves for the ground state of N$_2$.  (a) $\Lambda$-CI results.  (b) CI with a fixed number of determinants. (c) $\Lambda$-CI with a smoothed Hamiltonian [Eq.~\eqref{eq:Hsmooth}].  (d) $\Lambda$+SD-CI results. 
(e)-(f) $\Lambda$+SD-CI energy error with respect to FCI.
The curves obtained by correcting the $\Lambda$+SD-CI energy with the second-order estimate of the contribution from the discarded determinants are labeled $[E^{(2)}_{\rm corr}]$.
All computations employed the 6-31G basis set and restricted-Hartree--Fock orbitals.  The 1s--like orbitals of N were frozen in the computations of the correlation energy.}
\label{fig:n2}
\end{center}
\end{figure*}

We proceed to discuss the ground-state dissociation curve of N$_2$.
Results computed using a $\Lambda$-CI wave function and various values of $\Lambda$ are displayed in Fig.~\ref{fig:n2}-a.
For low values of $\Lambda$ (1--1.5 \Eh) the potential energy curve displays significant energy jumps, a consequence of the abrupt increase in size of the \Ml set.
For higher values of $\Lambda$, the $\Lambda$-CI the energy gaps become smaller, but irregularities in the potential energy curve can still be observed for the $\Lambda$ = 2.5 \Eh curve.
The best $\Lambda$-CI wave function considered here ($\Lambda =$ 3.5 \Eh) yields a curve that in the range 1--4 \AA{} deviates from the FCI curve at most by 26.1 m\Eh and has a nonparallelism error of 19.4 m\Eh.
For comparison, in the case of CAS(6,6)-CI the maximum error is 159 m\Eh and the NPE is equal to 43 m\Eh.

Since the change in size of \Ml caused energy jumps, it is interesting to study what happens when the energy-selected wave function contains a given number of determinants, which is fixed throughout the potential energy curve.
Fig.~\ref{fig:n2}-b shows the N$_2$ potential energy curve computed with wave functions that contain the energetically lowest 1000, 5000, and 25000 determinants.
These curves do not present large discontinuities, and their smoothness improves quickly as the size of the space is enlarged.
The potential energy curve obtained using 25000 determinants has a maximum error of 23.1 m\Eh and a NPE of only 14.2 m\Eh.
Nevertheless, fixing the size of the CI space is not a satisfactory solution because it destroys adaptivity.  This may be noticed for example, in the case of the wave function containing 5000 determinants: in the multireference limit the fraction of electron correlation recovered diminishes significantly.  

Another solution to the problem of discontinuities is to introduce a smooth truncation of the Hamiltonian matrix.
To this end, we introduce a second energy threshold $\Lambda_0$ and consider the following smoothed Hamiltonian $H^{\rm s}$:
\begin{equation} \label{eq:Hsmooth}
H_{IJ}^{\rm s} = H_{IJ} \times 
\begin{cases}
f_{\Lambda_0,\Lambda}(E_I) f_{\Lambda_0,\Lambda}(E_J) \;\; &I\neq J\\
1 \;\; &I = J
\end{cases}
,
\end{equation}
where $f_{\Lambda_0,\Lambda}(E)$ is the \textit{smootherstep}\cite{Ebert:2002:TMP:572337} function:
\begin{equation}
f_{\Lambda_0,\Lambda}(E) =
\begin{cases} 1 & E < \Lambda_0 \\
 6 t^5 - 15 t^4 + 10 t^3 \;\; & \Lambda_0 \leq E \leq \Lambda\\
0 & E > \Lambda
\end{cases},
\end{equation}
where the scaled energy $t$ is defined as:
\begin{equation}
t = \frac{\Lambda - E}{\Lambda - \Lambda_0}.
\end{equation}

$f_{\Lambda_0,\Lambda}(E) $ goes from 1 to 0 in the range $\Lambda_0 \leq E \leq \Lambda$ and has continuous first and second derivatives. 
The product $f_{\Lambda_0,\Lambda}(E_I) f_{\Lambda_0,\Lambda}(E_J)$ smoothly decouples determinants that fall in the energy range $\Lambda_0 < E < \Lambda$ by attenuating the off-diagonal elements of the Hamiltonian matrix.
Fig.~\ref{fig:n2}-c shows the energy computed by diagonalization of $H^{\rm s}$ in the case $\Lambda = 3$ \Eh and $\Lambda_0$ = 1 and 2.5 \Eh.
In both examples, it is found that the energy is a smooth function of the bond length.
For the $\Lambda_0$ = 2.5 \Eh curve, the maximum error is 63.6 m\Eh and the NPE is 30.5 m\Eh.
It is pleasing to see that the NPE for the smoothed $\Lambda$-CI wave function is smaller than that of the unsmoothed one with identical value of $\Lambda$ (36 m\Eh) and the NPE of the CAS(6,6)-CI wave function (43 m\Eh).
Smoothing the matrix elements of the Hamiltonian appears to be a viable solution to eliminate the discontinuity problem in the $\Lambda$-CI energy and generate a zeroth-order wave function of accuracy comparable to the CAS scheme.

Potential energy curves for N$_2$ computed with various $\Lambda$+SD-CI wave functions are shown in Fig.~\ref{fig:n2}-d.
All these use a very small reference $\Lambda$-CI wave function, obtained by setting $\Lambda$ = 1 or 2 \Eh and determinant selection based on $\epsilon_I^{(2)}$, setting $\tau_\epsilon$ = $10^{-6}$ and $10^{-9}$ \Eh.
For all the points sampled, the $\Lambda$-CI wave function for $\Lambda$ = 1 and 2 \Eh contains at most 204  and  3443 determinants, respectively, while the $\Lambda$+SD-CI wave function corresponding to $\tau_\epsilon$ = $10^{-9}$ \Eh contains at most 25785 and 76556 determinants, respectively.
Fig.~\ref{fig:n2}-d shows that on the scale of the dissociation energy of N$_2$, these small model spaces give $\Lambda$+SD-CI wave function that are significantly more accurate than the corresponding $\Lambda$-CI potential energy curves [compare with Fig.~\ref{fig:n2}-a].
The energy error with respect to FCI for the $\Lambda$+SD-CI curves based on the smaller reference space ($\Lambda$ = 1 \Eh) is shown in Fig.~\ref{fig:n2}-e.
For these wave functions, energy jumps up to about 10 m\Eh can be observed for $r$(N-N) less than 1 \AA{}, while at larger bond lengths the discontinuities are less pronounced.
Fig.~\ref{fig:n2}-f, shows instead that the $\Lambda$+SD-CI wave functions based on the $\Lambda$ = 2 \Eh model space are considerably more accurate and lead to smoother potential energy curves.
For the largest wave function in this series ($\tau_\epsilon$ = $10^{-9}$ \Eh) the nonparallelism error in the range 1--4 \AA{} is only 1.8 m\Eh (1.1 \kcal).
We also find that the second-order correction, $E^{(2)}_{\rm corr}$, added to the curves with $\tau_\epsilon$ = $10^{-6}$ and $10^{-7}$ \Eh [see Fig.~\ref{fig:n2}-f] reduces the error in $\Lambda$+SD-CI energy, but is not sufficient to match the accuracy of the uncorrected wave function with $\tau_\epsilon$ = $10^{-8}$ \Eh.

These examples show the versatility of the energy-based wave functions.
The $\Lambda$-CI scheme (in particular the smoothed variant) may be used to describe the zeroth-order wave function of a strongly-correlated system, while the $\Lambda$+SD-CI approach can be used to formulate more compact wave functions that can be applied to larger active spaces and used to efficiently and systematically approach the FCI energy.

\subsection{Low-lying excited states of C$_2$}
\begin{figure}[t]
\begin{center}
\includegraphics[width=3.5in]{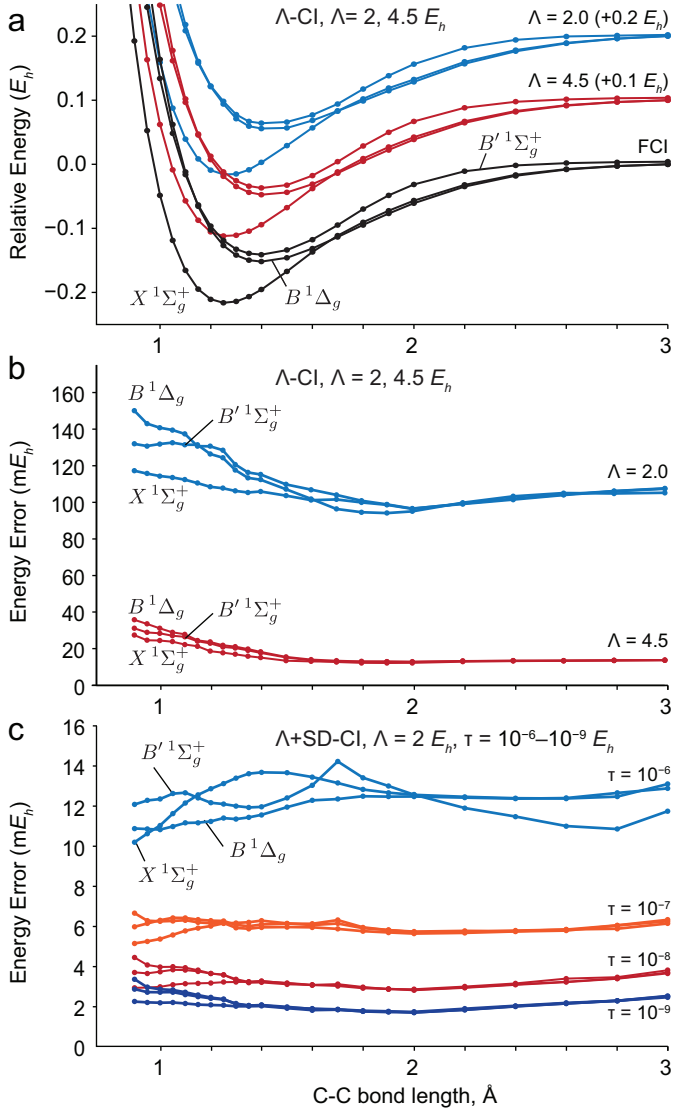}
\caption{$\Lambda$-CI and $\Lambda$+SD-CI potential energy curves for the $X\,{}^1\Sigma^{+}_g$ ground state and the $B\,{}^1\Delta_g$ and $B'\,{}^1\Sigma^{+}_g$ excited states of C$_2$.
(a) Total energy for FCI (from Ref.~\onlinecite{Abrams:2004ib}) and the $\Lambda$-CI wave function ($\Lambda$ = 2 and 4.5 \Eh).  As indicated in the plot, the $\Lambda$ = 2 and 4.5 \Eh curves were shifted by 0.2 and 0.1 \Eh, respectively.
(b) $\Lambda$-CI wave function, error with respect to the FCI curve.
(b) $\Lambda$+SD-CI wave function, error with respect to the FCI curve.
All computations employed the 6-31G* basis set and restricted-Hartree--Fock orbitals.
The 1s--like orbitals of C were frozen in the computations of the correlation energy.}
\label{fig:c2}
\end{center}
\end{figure}

To investigate the ability of the $\Lambda$-CI and $\Lambda$+SD-CI methods to describe near-degenerate electronic states we study the potential energy curve of the $X\,{}^1\Sigma^{+}_g$ ground state and the $B\,{}^1\Delta_g$ and $B'\,{}^1\Sigma^{+}_g$ excited states of C$_2$.
This molecule has been studied extensively with high-level \textit{ab initio} methods and it proves to be a quite challenging test case for new theories of electron correlation.\cite{Christiansen:1996ks,Leininger:1998de,Kowalski:2001ut,Piecuch:2002vx,Abrams:2004ib,Hirata:2004er,Piecuch:2004wv,Kowalski:2004cq,Sherrill:2005wg,Li:2006jo,Fang:2008ix,Karton:2009hw,Purwanto:2009gf,Booth:2011hg,Datta:2011es,Jiang:2011bw,Su:2011hh,Coe:2012hv,Cleland:2012fi,Datta:2012hu,Angeli:2012ih,Boschen:2014ed}
In Fig.~\ref{fig:c2}-a we report a comparison of the $\Lambda$-CI potential energy curves of C$_2$ and the FCI results of Abrams and Sherrill using the 6-31G* basis set and RHF orbitals.\cite{Abrams:2004ib}
Both the $\Lambda = 2$ \Eh and $\Lambda = 4.5$ \Eh curves can reproduce the qualitative features of the FCI potential, including the crossing of the $X\,{}^1\Sigma^{+}_g$ and $B\,{}^1\Delta_g$ states around 1.7\AA{} and the degeneracy of the three electronic states at the dissociation limit.
The error with respect to the FCI energy for the three electronic states studied with the  $\Lambda$-CI method is shown in Fig.~\ref{fig:c2}-b.
Interestingly, the energy error is almost unappreciable in the dissociation limit, and becomes larger as the C-C bond length approaches the equilibrium value and it is compressed.
However, the excitation energy from the $X\,{}^1\Sigma^{+}_g$ to the $B\,{}^1\Delta_g$ state computed near the FCI ground state equilibrium geometry ($r_{\rm C-C} = 1.25$ \AA{}) is already sufficiently accurate (for $\Lambda = 2 $ and $4.5$ \Eh, the error is respectively, 0.57 and 0.11 eV).

In Fig.~\ref{fig:c2}-c we show the results obtained for the $\Lambda$+SD-CI method based on a $\Lambda = 2$ \Eh reference wave function and $\epsilon_I^{(2)}$-based selection of the space $M_{\Lambda}^{\rm SD}$.
Notice that this plot is on a scale ten times smaller than that of Fig.~\ref{fig:c2}-b.
For values of $\tau_\epsilon = 10^{-6}$ \Eh, we already obtain potential energy curves with a nonparallelism error of the order of 4 m\Eh (ca. 0.06 eV).  More importantly, the maximum error in the excitation energy for the curves with $\tau_\epsilon = 10^{-7}$--$10^{-9}$ \Eh is found to be less than 2 m\Eh, which corresponds to ca. 0.03 eV.
This example highlights the ability of the $\Lambda$-CI and $\Lambda$+SD-CI wave functions to yield accurate excitation energies, even in difficult cases like that of C$_2$, which involves electronic states with a double excitation character.

\subsection{The bis-($\mu$-oxo) and $\mu$-$\eta^2$:$\eta^2$ peroxo forms of Cu$_2$O$_2^{2+}$}
In this section we apply the $\Lambda$-CI method to analyze the multireference nature of the Cu$_2$O$_2^{2+}$ core.
This system has been used by Cramer and co-workers,\cite{Cramer:2006jm,Cramer:2006ja} to model the active site of metalloenzymes that oxidize organic substrates via a copper-activated oxygen molecule and later by many others as a benchmark model for new theories.\cite{Azizi:2006iz,Kong:2008eo,Demel:2008ey,Malmqvist:2008eu,Marti:2008gz,Kurashige:2009gs,Gherman:2009ci,Saito:2010fo,Yanai:2010kf,Neese:2011hx,Shamasundar:2011ew,Liakos:2011kx,Samanta:2012fv} .
In this work we study the isomerization of the bare Cu$_2$O$_2^{2+}$ core from the bis-($\mu$-oxo) to the $\mu$-$\eta^2$:$\eta^2$ peroxo form.
Following the model of Cramer,\cite{Cramer:2006jm,Cramer:2006ja} the geometry is parameterized by the variable $F$ (ranging from 0 to 100) and the Cartesian coordinates of each atom $i$ ($\mathbf{q}_i$) are given by:
\begin{equation} \label{eq:cramer_model}
\mathbf{q}_i(F) = \mathbf{q}_i(0) + \frac{F}{100}[\mathbf{q}_i(100) - \mathbf{q}_i(0)],
\end{equation}
where $\mathbf{q}_i(0)$ and $\mathbf{q}_i(100)$ correspond respectively to the coordinates of the bis-($\mu$-oxo) and $\mu$-$\eta^2$:$\eta^2$ peroxo forms.

One of the most interesting findings of the study of Cramer and co-workers\cite{Cramer:2006ja} is that the Cu$_2$O$_2^{2+}$ systems are remarkably challenging for active space methods like CASSCF and CASPT2.
These methods yield isomerization energy curves that display substantial differences with respect to those computed using single-reference CC theory and the completely-renormalized CC approach employing the left eigenstates of the similarity-transformed Hamiltonian [CR-CCSD(T)$_{L}$].\cite{Piecuch:2005tl,Piecuch:2006fj}

We try to elucidate the nature of electron correlation in the Cu$_2$O$_2^{2+}$ system by computing the $\Lambda$-CI wave function for various values of $F$.
In our computation we set $\Lambda = 1$ \Eh and $\Lambda' = 4$ \Eh and use restricted Hartree--Fock orbitals computed with the cc-pVTZ basis set.  With these parameters, the size of the \Ml space is respectively, 16926 and 2578, for the bis-($\mu$-oxo) and $\mu$-$\eta^2$:$\eta^2$ forms, which correspond to a minute fraction ($10^{-14}$ and $1.5\times10^{-15}$) of the size of the FCI space, which contains about $1.7\times10^{18}$ elements.

\begin{figure*}[t]
\begin{center}
\includegraphics[width=6.5in]{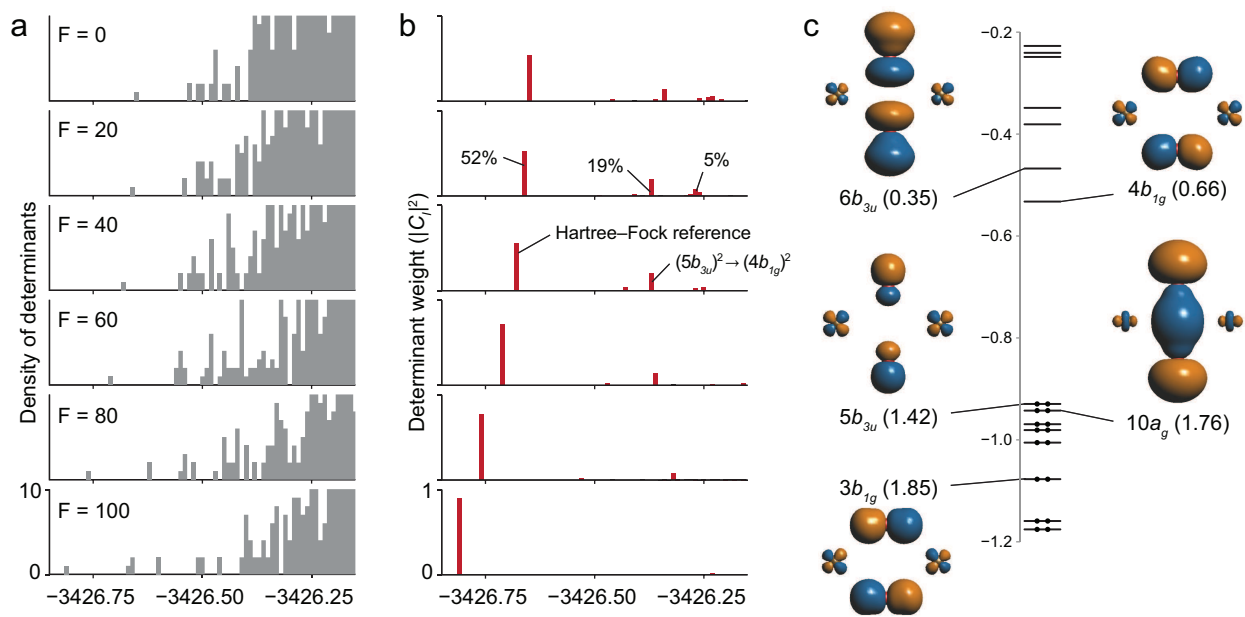}
\caption{Analysis of the $\Lambda$-CI/cc-pVTZ wave function ($\Lambda = 1$ \Eh, $\Lambda' = 4$ \Eh) of the Cu$_2$O$_2^{2+}$ system for selected values of $F$ in the range [0,100].  (a) Density of determinants for the $\Lambda$-CI wave function.  (b) Weight of the determinants in the $\Lambda$-CI wave function given by the square modulus of the corresponding coefficient ($|C_I|^2$). (c) Plot of the restricted Hartree--Fock orbitals and their energies (in \Eh) and the $\Lambda$-CI occupation numbers for the geometry corresponding to $F = 20$ (at isocontour density equal to 0.05).}
\label{fig:cu2o2}
\end{center}
\end{figure*}

Fig.~\ref{fig:cu2o2} illustrates the results of our analysis.
The leftmost panel, shows the density of determinants computed for various values of $F$.
In all cases, the Hartree--Fock reference is the determinant with the minimum energy.
Furthermore, the density of determinants is always \textit{gapped} (in the many-body sense), meaning that there is a sizable energy difference between the Hartree--Fock determinant and the lowest-lying excited determinant.

The center panel of Fig.~\ref{fig:cu2o2}, however, shows that at all geometries the wave function for this system is relatively simple.
For values of $F \leq 40$, the $\Psi_{\Lambda\text{-CI}}$ is significantly multiconfigurational but only a few determinants have a significant weight.  For example, if we look closer at case $F=20$, we notice that the wave function is dominated by the Hartree--Fock determinant (52\% contribution to $\Psi_{\Lambda\text{-CI}}$) and there are only two additional large contributions from the the $(5b_{3u})^2 \rightarrow (4b_{1g})^2$ and $(10a_{g})^2 \rightarrow (6b_{3u})^2$ double excitations.
The weight of these doubly-excited determinants is 19\% and 5\%, respectively, while the rest of the determinants have a weight of about 2\% or less.
The rightmost panel of Fig.~\ref{fig:cu2o2} displays the molecular orbital diagram for the $F=20$ geometry.
There is a large energy gap between the HOMO and LUMO, but interestingly, the occupation numbers for the $10a_g$, $5b_{3u}$, $4b_{1g}$, and $6b_{3u}$ orbitals deviate significantly from the Hartree--Fock reference.

\begin{figure}[t]
\begin{center}
\includegraphics[width=3.5in]{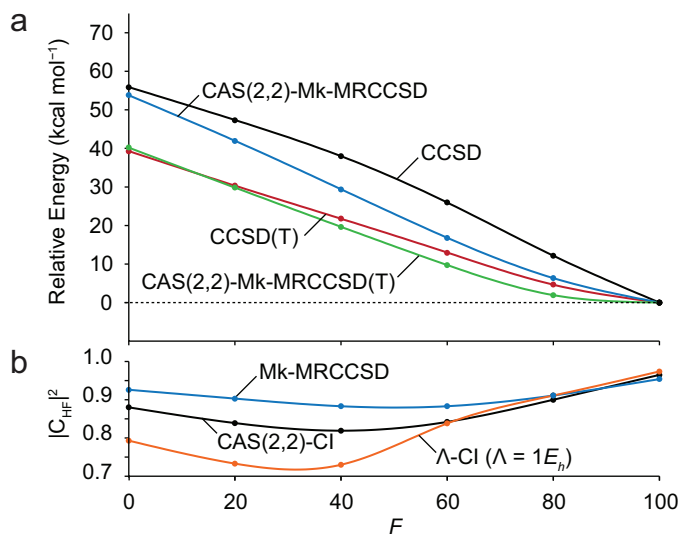}
\caption{Cu$_2$O$_2^{2+}$ model system.  (a) Relative energy with respect to the $\mu$-$\eta^2$:$\eta^2$ peroxo form along the reaction coordinate specified by Eq.~\eqref{eq:cramer_model}.  (b) Weight of the Hartree--Fock reference computed using the CAS(2,2)-Mk-MRCCSD approach, CAS(2,2)-CI, and $\Lambda$-CI wave functions.  All computations, including those employing the CAS(2,2)-Mk-MRCCSD method, used the cc-pVTZ basis set and restricted-Hartree--Fock orbitals.}
\label{fig:cu2o2_mkmrcc}
\end{center}
\end{figure}

This analysis can help the selection of an appropriate active space for a more elaborate multireference computation.
This point is illustrated by a state-specific multireference coupled cluster computation using the Mukherjee method (Mk-MRCC).\cite{Mahapatra:1999tm}
We report results using Mk-MRCC with the singles and doubles approximation (Mk-MRCCSD)\cite{Evangelista:2007hz} and the Mk-MRCCSD method with perturbative triples [Mk-MRCCSD(T)], as implemented in Ref.~\onlinecite{Evangelista:2010cq}.
Because the $(5b_{3u})^2 \rightarrow (4b_{1g})^2$ double excitation is the second-largest contribution to the $\Lambda$-CI wave function, it appears reasonable to use a CAS(2,2) generated by distributing two electrons in the $5b_{3u}$ and $4b_{1g}$ orbitals.
Fig.~\ref{fig:cu2o2_mkmrcc} shows the relative energy of the Cu$_2$O$_2^{2+}$ as a function of $F$, with the $F=100$ geometry taken as reference.
The relative energy at $F = 0$ for the Mk-MRCCSD and Mk-MRCCSD(T) methods is respectively, 53.8 and 40.2 \kcal.
Single-reference CC methods yield very similar results, 55.9 and 39.3 \kcal respectively for the CCSD and CCSD(T) approaches.
The Mk-MRCCSD(T) and CCSD(T) results are in good agreement with all the CR-CCSD(T)$_{L}$ results reported in Ref.~\onlinecite{Cramer:2006ja}, which use effective-core potentials and a different basis set.  For example, the CR-CCSD(T)$_{L}$ isomerization energy is 35.5 \kcal, and when quadruples are included [CR-CCSD(TQ)$_{L}$], it increases slightly to 38.5 \kcal.
Interestingly, the isomerization energy computed with the $\Lambda$-CI wave function has the wrong sign (ca. $-25$ \kcal), a behavior also displayed by projected Hartree--Fock wave functions.\cite{Samanta:2012fv}
The CASSCF(16,14) and CASPT2(16,14) results\cite{Cramer:2006jm} (0.2 and 7.2 \kcal, respectively) also differ significantly from the CC results.

The lower panel of Fig.~\ref{fig:cu2o2_mkmrcc} also shows the importance of the Hartree--Fock configuration in the Mk-MRCCSD wave function.
In agreement with the $\Lambda$-CI analysis, the maximum multireference character is displayed at geometries that are intermediate between the two extremes.
In the Mk-MRCC zeroth-order wave function, the Hartree--Fock reference accounts for a large part of the state vector (88\% or more), while in a CAS-CI(2,2) and in the $\Lambda$-CI wave functions the double excitation $(5b_{3u})^2 \rightarrow (4b_{1g})^2$ is given more importance.

The agreement between the single- and multireference CC results and the predominance of the Hartree--Fock determinant in the Mk-MRCC wave function suggest that the Cu$_2$O$_2^{2+}$ core is at best a mild multireference problem.
Thus, it appears that the failure of the CASPT2 method should not be ascribed to strong correlation effects, but to the fact that in the Cu$_2$O$_2^{2+}$ system \textit{dynamical electron correlation} must be treated beyond the perturbative regime.
This finding is in agreement with earlier conclusions of Cramer \textit{et al.}\cite{Cramer:2006jm,Cramer:2006ja} and Neese and co-workers.\cite{Neese:2011hx,Liakos:2011kx}

\section{Energy separability and orbital invariance properties of the $\Lambda$-CI approach}
In this section we discuss some of the formal properties (or lack thereof) of the $\Lambda$-CI wave function.
As for truncated CI methods, the $\Lambda$-CI energy of noninteracting fragments is not equal to the sum of the individual fragment energies.
However, due to its adaptive nature, by increasing $\Lambda$ the separability error can be made arbitrarily small.
We demonstrate this point by considering a system of noninteracting helium atoms.
Table~\ref{tab:size-consistency} shows the separability error, for clusters containing from two to four He atoms.  For small values of $\Lambda$, the separability error is null because the CI space contains single excitations that according to Brillouin's theorem have zero coupling with the Hartree--Fock determinant.
For larger values of $\Lambda$ (4--6 \Eh) the separability error has an erratic behavior, likely caused by the uneven inclusion of determinants in the monomer and clusters.  However, for values of $\Lambda \geq 7$ \Eh the separability error assumes a regular behavior and decreases monotonically with respect to $\Lambda$.

The second property that we consider is the invariance of the energy with respect to rotations of the orbitals.
First, we note that in traditional wave function methods that rely on an orbital partition, it is natural to ask whether or not the energy is invariant with respect to rotations of orbitals within subspaces.
On the contrary, in the case of the $\Lambda$-CI approach, the wave function is not defined by a partition of the orbitals and therefore the concept of orbital space must emerge from the property of the wave function itself.
This can be done, once the $\Lambda$-CI wave function is built, by computing the occupation numbers of each orbital:
\begin{equation} \label{eq:occupation_number}
n_p = \bra{\Psi_{\Lambda\text{-CI}}} \cop{p} \aop{p}\ket{\Psi_{\Lambda\text{-CI}}},
\end{equation}
and defining classes of orbitals characterized by having equal occupation numers.
For sufficiently small values of $\Lambda$, two important classes of orbitals might arise in this analysis: those with $n_p$ equal to zero or one.
These two classes of orbitals identify orbitals that are fully occupied (core) and unoccupied (virtual) orbitals in every determinant that is contained in \Ml.
Since the energy of a determinant is invariant with respect to separate rotations of the orbitals that enter its definition---and trivially also those that do not---then rotating orbitals within the occupied or unoccupied class will leave the $\Lambda$-CI wave function and energy invariant.
In the case of partially occupied (active) orbitals ($ 0 < n_p < 1$) the invariance property is generally lost.
This situation is similar to the case of a general multiconfigurational wave function, for which a rotation of the partially occupied orbitals change the wave function and energy.
\cite{Evangelista:2011ij}

\begin{table*}[htbp]
\centering
\caption{Separability error of the $\Lambda$-CI energy (expressed in m$E_h$) computed for clusters of noninteracting helium atoms using the cc-pVDZ basis set and restricted-Hartree--Fock orbitals.}
\label{tab:size-consistency}
\begin{tabular*}{6.25in}{@{\extracolsep{\fill}}cccccccccc} 
\hline\hline
& \multicolumn{8}{c}{$\Lambda$ (\Eh)}\\
Separability error (m$E_h$) & 2 & 3 & 4 & 5 & 6 & 7 & 8 & 9 & 10 \\
\hline
$E({\rm He}_2) - 2 E({\rm He})$ &   0.00 &    0.00 &   29.99 &    0.13 &   35.00 &    0.34 &    0.34 &    0.34 &    0.11\\
$E({\rm He}_3) - 3 E({\rm He})$ &   0.00 &    0.00 &   29.99 &    0.38 &   35.49 &    1.25 &    1.02 &    1.02 &    0.32\\
$E({\rm He}_4) - 4 E({\rm He})$ &   0.00 &    0.00 &   59.99 &    0.76 &   70.51 &    2.47 &    2.47 &    2.01 &    1.56\\
\hline\hline
\end{tabular*}
\end{table*}

\section{Discussion}
We have presented a methodology to construct zeroth-order model spaces and wave functions  based on a simple criterion: our scheme selects only those determinants that lie within a given energy threshold ($\Lambda$) from the lowest energy determinant.
By screening the determinants according to their energy,  the $\Lambda$-CI approach yields a model space that is adaptive and  does not require the selection of a set of active orbitals.
In addition, the $\Lambda$-CI approach is systematically improvable, as for $\Lambda\rightarrow \infty$, the wave function converges to the FCI limit.
This implies that the separability and orbital invariance errors inherent to this method can be made arbitrarily small.
In addition, we introduced a selected multireference CISD wave function based on the $\Lambda$-CI model space ($\Lambda$+SD-CI).
In the $\Lambda$+SD-CI approach the singly and doubly excited determinants generated from the $\Lambda$-CI model space are screened according to an importance criterion.
For CI spaces of the same dimensions, the $\Lambda$+SD-CI wave function recovers a larger fraction of the correlation energy than the $\Lambda$-CI scheme.

Our results show that these adaptive wave functions are versatile: the $\Lambda$-CI method can be used to generate compact zeroth-order model spaces, and the $\Lambda$+SD-CI scheme offers an efficient way to deal with larger active spaces.   These points were illustrated in our computations of the ground state dissociation curve of N$_2$ and the potential energy curve for the first three singlet electronic states of C$_2$.
Our last example, which involves the Cu$_2$O$_2^{2+}$ system, illustrates the use of the $\Lambda$-CI approach to diagnose the multireference character of an electronic state and design an appropriate active space for a subsequent multireference CC computation.

From the computational point of view, both the $\Lambda$-CI and $\Lambda$+SD-CI methods have the advantage of being strikingly simple and computationally robust: unlike the case of CASSCF and DMRG methods, which perform a nonlinear optimization, the adaptive schemes are noniterative and require only matrix diagonalization.
All the results presented in this paper were obtained with a pilot implementation.  We expect that the performance of these approaches can be greatly improved by using a sparse CI vectorized algorithm that does not store the Hamiltonian matrix.\cite{Knowles:1989te,Rolik:2008ik}
This will allow computations with several hundred million determinants.
In addition, when the $\Lambda$-CI wave function is used as a diagnostic tool, its computational cost is significantly smaller than that required to perform a CASSCF computation.  Thus a $\Lambda$-CI computation could be in principle performed routinely after a mean-field computation and warn the user of a potential multireference problem.

The $\Lambda$-CI  and $\Lambda$+SD-CI methods rely on the fundamental assumption that the nondynamical correlation that enters the problem under consideration is a local (size intensive) effect.
In other words, we postulate that even if the size of the Hilbert space grows factorially, the space of important configurations remains small and we can thus pick the elements that belong to it.
This is perhaps the only scenario in which it makes sense to adaptively select a zeroth-order wave function.
When nondynamical correlation will grow with the size of the system, like for example in the case of a periodic lattice of transition metal oxides, then the $\Lambda$-CI approach will fail.
In this case, if the wave function can be factorized, then approaches like DMRG will provide the optimal solution.

The wave functions proposed in this work are meant to describe the static component of electron correlation, and to be used in applications that require active spaces that go beyond the current limits of the CASSCF approach.
Although these adaptive wave functions can be used to achieve the accuracy required to describe ground state thermochemistry and electronic excited states, it is more convenient to combine them with an approach that can efficiently treat the dynamic component of electron correlation.
We think that the most promising way to achieve this goal is within an equation-of-motion formalism.
There are several ideas worth exploring.
For example, it is well appreciated\cite{Bartlett:2007kv} that excited states with a large component of double excitation character require introducing expensive triple excitations in the EOM-CC formalism.
It would be interesting to modify the single-reference EOM-CC formalism to use the adaptive basis of excited configurations \Ml  to diagonalize the similarity-transformed Hamiltonian.
We expect that this adaptive EOM-CC approach would be able to accurately describe electronic states with different excitation character without resorting to a full treatment of triple excitations.
Another attractive idea, is to use the adaptive model space in a multireference EOM-CC formalism.\cite{Krylov:2008ud,Musiai:2011dw,Datta:2012hu}
These are all topics that our laboratory will explore in the future.

The development of adaptive electronic structure methods presents clear challenges and requires a significant paradigm shift.
In order to gain robustness, versatility, and control over the accuracy of a computation it might be worth compromising certain formal properties like size extensively, energy separability, and orbital invariance.
If we abandon these strict requirements, we are left with a vast number of exciting possibilities to explore.

\begin{acknowledgments}
This work was supported by start-up funds provided by Emory University.
\end{acknowledgments}

\appendix
\section{An alternative algorithm for building \Ml}
The second algorithm that we present builds \Ml using a breadth-first search in excitation space coupled with pruning.
We start from a reference determinant $\Phiref$ with corresponding energy $\tilde{E}$ that is reasonably close to $E_0$.  
Accordingly, the minimum energy is initialized to $E_{\rm min} = \tilde{E}$.
From $\Phiref$ we generate all the singly excited determinants $\tilde{\Phi}_i^a = \cop{a}\aop{i} \Phiref$, where the indices $i,j,\cdots$ and $a,b,\cdots$ refer respectively to occupied and virtual orbitals of the Fermi vacuum $\Phiref$.
The relative energy of the determinant $\tilde{\Phi}_i^a$ can be easily computed as [see Eq.~\eqref{eq:det_energy_difference}]:
\begin{equation}
\label{eq:fast_single}
E_i^a - \tilde{E} = \bra[1]{\tilde{\Phi}_i^a} \hat{H} \ket[1]{\tilde{\Phi}_i^a} -\tilde{E} = \epsilon_{a}(\Phiref) - \epsilon_{i}(\Phiref) - V_{ia},
\end{equation}
where $ \epsilon_p(\Phiref)$ is given by Eq.~\eqref{eq:epsilon}.
From the list of singly excited determinants we select those with relative energy $E_i^a-E_{\rm min}<\Lambda'$ (with $\Lambda' \geq \Lambda$) and collect them in the set $M^{(1)}_{\Lambda'}$.
At the same time we also update the value of $E_{\rm min}$ if a lower determinant energy is found.

For each determinant $\tilde{\Phi}_i^a \in M^{(1)}_{\Lambda'}$ we then proceed to generate doubly-excited determinants of the form $\tilde{\Phi}_{ij}^{ab}$ with $i > j$ and $a < b$.
This restriction is imposed to avoid generating the doubly excited determinant $\tilde{\Phi}_{ij}^{ab}$ from two different singly-excited determinants, for example from $\tilde{\Phi}_{i}^{a}$ via the excitation $j\rightarrow b$ or from $\tilde{\Phi}_{j}^{a}$ via the excitation $i\rightarrow b$.
The relative energy of the determinant $\tilde{\Phi}_{ij}^{ab}$ may be evaluated using the equation
\begin{equation}
\label{eq:fast_double}
E_{ij}^{ab} - E_{i}^{a} = \epsilon_{b}(\tilde{\Phi}_{i}^{a}) - \epsilon_{j}(\tilde{\Phi}_{i}^{a}) - V_{jb},
\end{equation}
which is a generalization of Eq.~\eqref{eq:fast_single}.
Notice that the orbital energies $\epsilon_{p}(\tilde{\Phi}_{i}^{a})$ that enter Eq.~\eqref{eq:fast_double} are computed using Eq.~\eqref{eq:epsilon} from the occupation numbers of $\tilde{\Phi}_{i}^{a}$.
The doubly-excited determinant that satisfy $E_{ij}^{ab}-E_{\rm min}<\Lambda'$ are then included in the set $M^{(2)}_{\Lambda'}$.

This process is repeated until we reach an excitation level $k_{\rm max}$ such that $M^{(k_{\rm max} + 1)}_{\Lambda'}$ is empty.
At the end, the set \Ml is given by the union of all the sets $M^{(k)}_{\Lambda'}$ excluding those elements that have a relative energy greater than $\Lambda$:
\begin{equation}
M_\Lambda = \left\{\Phi_I \in \bigcup_{k=1}^{k_{\rm max}} M^{(k)}_{\Lambda'} : E_I - E_0 \leq \Lambda \right\}.
\end{equation}

\newpage


\begin{thebibliography}{139}
\bibitem [{1()}]{1}%
  \BibitemOpen
  \href@noop {} {}\bibinfo {note} {In the case of the FCI wave function,
  Gershgorin's circle theorem can be used to interpret the density of
  determinants as an approximation of the exact eigenvalue spectrum of the
  Hamiltonian.}\BibitemShut {Stop}%
\bibitem [{3()}]{3}%
  \BibitemOpen
  \href@noop {} {}\bibinfo {note} {The nonparallelism error (NPE) is commonly
  defined as the difference between the maximum and minim of the error with
  respect to FCI [$\Delta E(r)$] over a range of geometries $R$: $\mathrm {NPE}
  = \max _{r \in R}[\Delta E(r)] - \min _{r \in R}[\Delta E(r)]$}\BibitemShut
  {NoStop}%
\end{thebibliography}
%

\end{document}